\documentclass[11pt,twoside,A4]{article} 
\usepackage{times,fancyhdr}
\usepackage[dvips]{graphicx}
\usepackage{latexsym} 
\usepackage[affil-it]{authblk}
\usepackage{amsmath}
\usepackage{amssymb}
\usepackage{setspace}

\usepackage[top=4cm, bottom=4cm, left=4cm, right=4cm]{geometry}

\def\beq{\begin{equation}}
\def\eeq{\end{equation}}
\def\beqn{\begin{eqnarray}}
\def\eeqn{\end{eqnarray}}

\begin{document}
 
\title{The remote Maxwell demon as energy down-converter}
\author{Sabine Hossenfelder\thanks{hossi@nordita.org}} 
\affil{\small Nordita\\
KTH Royal Institute of Technology and Stockholm University\\
Roslagstullsbacken 23, SE-106 91 Stockholm, Sweden}
\date{}
\maketitle
\begin{abstract}
It is demonstrated that Maxwell's demon can be used to allow a machine to extract energy from a heat bath by use of information that is processed by the demon at a remote location. The model proposed here effectively replaces
transmission of energy by transmission of information. For that we use a feedback protocol that enables a
net gain by stimulating emission in selected fluctuations around thermal equilibrium. We estimate
the down conversion rate and the efficiency of energy extraction from the heat bath.
\end{abstract}

\section{Introduction}

Maxwell's demon \cite{Maxwell,Sz} uses information to do work. It is by now understood
that to store, manipulate and erase
information the demon either has to give up its own low entropy state,
or it must have access to an energy source \cite{Landauer, Bennett,cost01,cost02,demon02,demon03}. The demon's function thus, while
paradoxical at first sight, is in perfect agreement with the laws of thermodynamics,
but  it held and continues to hold many lessons about the relation between information and energy.

Far from being a mere thought experiment, by now Maxwell's demon can and
has been realized in the laboratory. It has been demonstrated that information can be converted into
work indeed and Maxwell's demon has become reality \cite{demon,demon1}. 
The
study of these systems has been very fruitful for the understanding of non-equilibrium
statistical mechanics and the energy cost of information. 
More recently, the quantum version
of Maxwell's demon has spurred interest \cite{qdemon}, which highlights the relation between
statistical mechanics, thermodynamics and quantum information \cite{qi}, research that bears
relevance not only for quantum computing but also for quantum gravity.

Here, we are not  concerned with the quantum version, but focus on the classical Maxwell
demon from a new perspective, that of the local 
flow of energy. 

There is, in principle, no reason why the demon has to do its work  the system that eventually generates work from the demon's information,
hereafter referred to as `the machine'. The
machine can take energy from a heat bath it is embedded in, but without
information this energy is not useful and cannot be exploited to do work. However, this
opens up the possibility that the demon can work remotely, powered by some source of
energy, and transmit to the machine only the information that allows to extract energy from the
heat bath.

A similar idea was exploited
for the recently proposed quantum energy teleportation \cite{Hotta:2008uk, Hotta:2013qta}, which
uses entanglement of the quantum vacuum. Since we are not dealing with
quantum information but with classical information, we cannot, one the one hand, exploit entanglement. On the
other hand, there is nothing that prevents us from copying classical information so it can be
processed elsewhere. The system of the remote demon and the machine then essentially work as
an energy down-converter: Instead of transmitting energy directly, only the information
necessary to extract the energy is transmitted.

We will here lay out a model for how such a
system can operate and estimate its efficiency.

\section{Model for the remote demon}

The machine is placed
in a heat bath with temperature $T$. The demon has an energy-reservoir and is not
in equilibrium with the heat bath; it has its own heat bath at temperature $T_{\rm D} < T$.  We are in the following not interested to construct
a demon able to efficiently use the heat bath. Instead, we are interested in using the
demon to allow the machine to efficiently use the heat bath by means of a feedback protocol 
exploiting statistical fluctuations, similar in spirit to the protocol used in \cite{demon}. 

\begin{figure}[ht]
\centering 
\includegraphics[width=12cm]{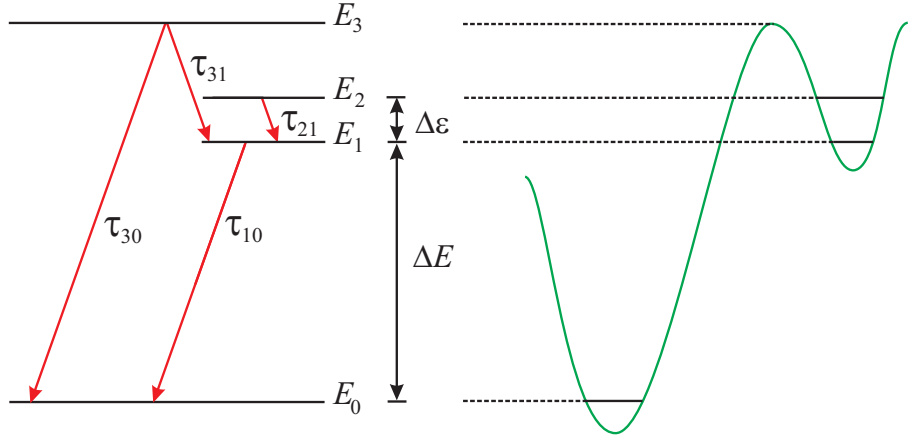} 

\caption{Nomenclature of energy levels and decay times of the machine's cells. \label{potential}}
\end{figure}

In order to keep track of the
flow of energy in the system, we use a four-level system ( figure \ref{potential}, left). The
machine is composed of elements that each constitute such a four-level system.

In the four-level system,
the ground level is $E_0$, and $E_1$ is a long-lived state  that only slowly decays
to $E_0$ with decay time $\tau_{10} \gg 1/T$. The level $E_1$ can be excited
into the level $E_2$ which decays very quickly into $E_1$ with decay time
$\tau_{21} \ll \tau_{10}$.  The fourth and highest level $E_4$ can decay into
$E_0$ or $E_1$. We denote the energy differences as
$E_1-E_0 = \Delta E$ and $E_2-E_1 = \Delta \varepsilon$,
where we assume $\Delta E \gg \Delta \varepsilon$.

This setting is suggestive of atomic energy levels, but
the model depends only on the general level structure. Another
way to realize such a four-level scheme is when the upper level $E_3$ is unstable 
and represents instead a potential barrier between the ground state $E_0$ and the
levels $E_2$ and $E_3$ ( figure \ref{potential}, right). These two scenarios have the common 
properties that the number of states in $E_1$ can be measured with an energy
smaller than $E_1-E_0$. Note that it has recently
become technologically feasible to create `artificial atoms' with custom designed 
potential levels that could find an application here \cite{artificialatoms}.

The purpose of the remote demon is to extract energy from the machine's heat bath by
use of information. This information can only be in deviations from thermal
equilibrium; we will here use temperature fluctuations. To make use of the
thermal fluctuations, we
divide the machine up into $N$ cells small enough so that statistical deviations
from thermal equilibrium can be detected. Each cell contains $n$ of the elements
that each are one of the four-level systems, so the demon consists of $Nn$ elements. 
Since convergence to the mean
goes with $1/n$, ie the average fluctuations decrease the larger the size
of the cell, we will want the cells to be as small as possible. 

We will assume that the exchange of information and
energy between
{\bf D} and {\bf M} is very fast and the signal extremely unlikely to interact with the heat bath, thus
not in thermal equilibrium.

\subsection{Simplified Case}

We will first discuss a single instance of the case with $n=1$, then turn to the
cyclic case with arbitrary $n$. The level configuration is akin that of the 3-level laser, in which stimulated emission
can take place between the levels $E_1$ and $E_0$ at the frequency $\Delta E$, just that we
have an extra resonance added between $E_1$ and $E_2$. Suppose then we
measure whether a state is in $E_1$ by use of the resonance between $E_1$ and
$E_2$ at energy $\Delta \varepsilon$, and then induce emission of the excited states with energy $\Delta E$. In this
case we will have invested the energy $\Delta E + \Delta \varepsilon$ but are returned the energy $2 \Delta E$. 
This emitted energy can be measured as  pressure and
does work $W=2\Delta E$. We assume that the energy $\Delta \varepsilon$ is thermally reemitted
and heats up the bath.

The reason we can extract energy from the heat bath in this process is that we made use of information about
which states are excited. Without that information, we would be more likely to lose energy into
absorption as to stimulate emission, thus not getting a net gain. 

Let us define the energy conversion factor $h = W/(W_1 + W_2)$ as the ratio of the
work extracted by the machine over the energy obtained by the machine from the demon.  In the case $n=1$, we have then $h =2 /(1+ \epsilon)$, where $\epsilon = \Delta \varepsilon/\Delta E$. That is provided that the
probability of the signal to cause stimulated emission is equal to one. In the general case, it
will be $h = (1+ \sigma_{10})/(1+\epsilon) $, where $\sigma_{10}$ denotes the probability of the
demon's signal to stimulate the emission from $E_1$ to $E_0$.

\subsection{General Case}

\begin{figure}[ht]

\includegraphics[width=14cm]{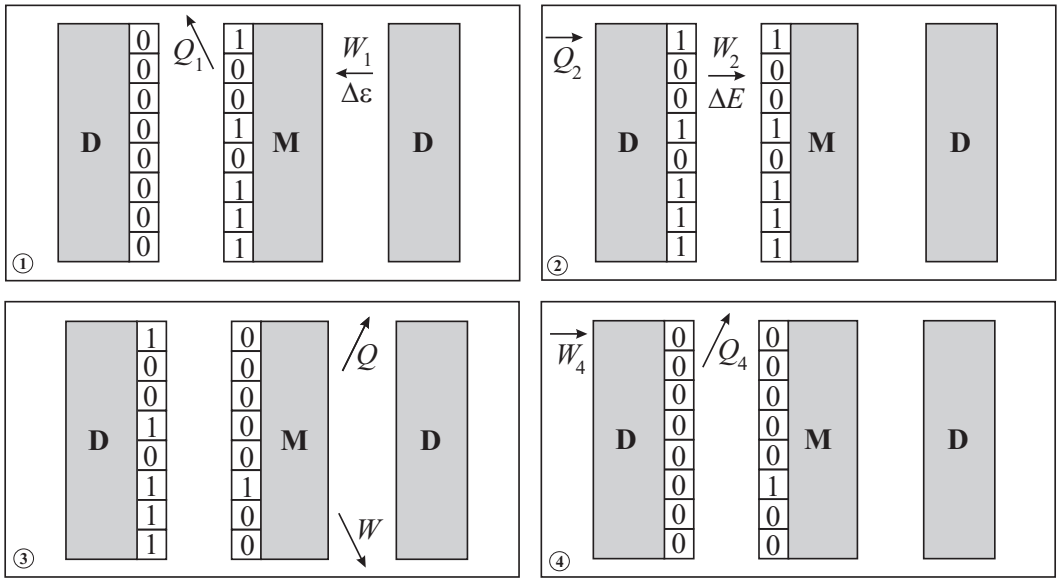}

\caption{Flow of work and energy for the general case of the remote Maxwell demon. Description see text. 
  For $n>1$ we can imagine the elements of the cells being lined up in the direction perpendicular to the
image. The part of the demon that is left in the figure would then be located on the perpendicular axis too.
This 3-dimensional spatial arrangement 
is not shown in the figure for graphical reasons.
 \label{fig2}}
\end{figure}

In the general case the demon, {\bf D}, and the machine, {\bf M}, operate in series of four steps illustrated in figure \ref{fig2}. In the
figure the demon is composed of two parts, but this is merely to make the process better to illustrate. The
demon has access to an energy source and is coupled to the heat bath at temperature $T_{\rm D}$. The
boxes with numbers in the machine are the cells of the machine. We label them with $1$ if all elements of the cell are excited and with $0$ otherwise. 

\begin{itemize}
\item[] {\bf Step 1 (measure)}: The demon at $T_{\rm D}$ is in a low entropy state, shown as
series of $0$s in Figure \ref{fig2}. The machine is in thermal equilibrium with its heat bath and has a thermal
distribution of excited states, shown as $1$s and $0$s. The right part of the demon emits the measuring
signal of energy $\Delta \varepsilon$, the resonance energy of the excited states of {\bf M}. 
To emit this signal, the demon needs an energy $W_1 \geq N n \Delta \varepsilon$, which is partly
absorbed and reemitted by {\bf M} into the heat bath as $Q_1$.

\item[] {\bf Step 2 (copy and compute):} The signal not absorbed or scattered  is detected
at the left side of the demon. Since the resonant state is short-lived compared to
the excited state $E_{1}$, 
this measurement is unlikely to kick the excited state into the ground state. This means 
that the measurement effectively copies the information from {\bf M} to {\bf D}.

The demon now 
sends a signal of energy $\Delta E$ to the cells with all excited elements, using energy $W_2 \geq \Delta E$ each. For this the demon has to first condense the information about $n$ elements per cell. It will
in general not succeed in stimulating emission for all cells in state $1$, both
because the probability for inducing emission is smaller than 1 and because
the demon, if not at zero temperature, will make errors due to thermal fluctuations.
This is indicated in the figure as a remaining 1 in the machine.

\item[] {\bf Step 3 (cool)}: Since
the demon targets only cells that are at energies above the statistical average, it has an above
average probability of inducing emission, for estimates see next section. In that
process the machine extracts energy $Q$ from the heat bath by use of the information the demon
has gathered and produces work $W$.

\item[] {\bf Step 4 (erase)}: The demon  erases the information it had
copied from the machine by using work $W_4$, and returns to its initial state and. The machine returns 
to equilibrium with the heat bath and the demon can resume with Step 1.
\end{itemize}

\section{Estimated Efficiency}

In thermal equilibrium, each level (${\rm i} \in \{1,2,3,4\}$) of the machine's cells will
be occupied with probability 
\beqn
p^{\rm M}_{\rm i} = \exp\left(- \frac{E_{\rm i}}{T} \right) / Z^{\rm M}\quad, 
\eeqn
with the normalization
\beqn
Z^{\rm M} = \sum_{\rm{i}} \exp \left(- \frac{E_{\rm i}}{T}  \right) \quad.
\eeqn
The average number of machine elements in state ${\rm i}$ is $n N p^{\rm M}_{{\rm i}}$ and the
average number of cells with $n$ elements that contain only excited elements is $N (p^{\rm M}_1)^n$. Let us
first assume that $T_{\rm D} = 0$.

If the demon targets all excited cells correctly, it will send an average energy of $W_2= \Delta E N (p^{\rm M}_1)^n$
and the average energy gained by the machine will be $W = \Delta E (n+1) N (p^{\rm M}_1)^n$. The energy down-conversion
factor is thus 
\beqn
h_n=\frac{1+n}{1+\epsilon_n} ~,~ \epsilon_n = (p_1^{\rm M})^{-n} \frac{\Delta \varepsilon}{\Delta E}~.
\eeqn 
The larger we make $h_n$ by increasing $n$, the less work the
machine will in total generate because the probability of finding a cell with all excited elements decreases
exponentially with $n$. Or, to put it differently, we would have to make the machine exponentially larger
to keep the total work the same with a larger conversion factor.

If we take into account the finite probability $\sigma_{10}$ of the demon's signal to stimulate 
emission, we find by counting the possibilities of failure to induce emission in any combinations of
the cell's elements that
\beqn
W &=&  \Delta E N (p_{1}^{\rm M})^n q_n~,~ h_n = \frac{1}{1+\epsilon_n} q_n~,
\eeqn
where
\beqn
q_n := \sum_{k=0}^n (1+k) {n \choose k} \sigma_{10}^k (1-\sigma_{10})^{n-k}~.
\eeqn

The possible distance between {\bf D} and {\bf M} is limited by the lifetime of $E_1$. We can take it into account by replacing
$\sigma_{10}$ with $\sigma_{10}(t) = \sigma_{10} \exp(-\Delta t/\tau_{10})$. If the demon needs much longer to send its signal than the lifetime of the state, then the possible energy gain will deteriorate.

We define the efficiency of the machine as
\beqn
\eta_{\rm M} = \frac{W-W_2-Q_1-Q_4}{Q} \quad.
\eeqn
That is, we subtract from the work $W$ that the machine generates the work that the demon sent
as input, and take into account that $Q_1$ and $Q_4$ are deposited into the heat bath to erase the
information of the measurement.

The 4-step
procedure does not pin down exactly which part of the heat the demon deposits into the bath at each step.
We will assume that the machine absorbs part of the measuring signal $W_1$ that came
from the demon and returns it as heat to the bath. The remainder of the signal, which is not absorbed
and serves to copy the machine's state, is removed by the demon in step $4$. There is thus no
work done with $Q_1+Q_4$. The energy $Q$ that is extracted from the heat bath is the energy necessary to refill the depleted
energy levels of the machine, ie $Q=W-W_2$.

At $T_{\rm D} > 0$ the demon has a non-vanishing probability to 
falsely send a signal to a cell that is not excited, or to fail to send a signal to an excited cell.
This can be estimated considering the demon's information-processing
part (left side in figure \ref{fig2}) to be composed of $N$ two-level system -- one
per each element of the machine -- with an
energy difference $\Delta E$, like the cells of the machine.   

At a temperature of
$T_{\rm D}$ the probability for an accidental emission is typically $p_1^{\rm D}:=v/(1+v)$ with $v=\exp(-\Delta E/T_{\rm D})$, ie if the temperature of the demon is considerably higher than the energy of the signal it sends, it will be error-prone. 

With the noise added, the demon sends a signal
\beqn
W_2 = N \Delta E (p_1^{\rm M})^n p_0^{\rm D} + N\Delta E p_1^{\rm D} (1 - (p^{\rm M}_1)^n)~, 
\eeqn
where $p_0^{\rm D}=1-p_1^{\rm D}=1/(1+v)$.
Here, the first term is the desired signal reduced by the demon's failure to respond because there
was nothing in the lower level, and the second term is faulty emission due to thermal excitation.
The first term thus decreases the work that the machine can do, while the latter increases its heat
because the signal is absorbed. 

\begin{figure}[ht!]

\includegraphics[width=14cm]{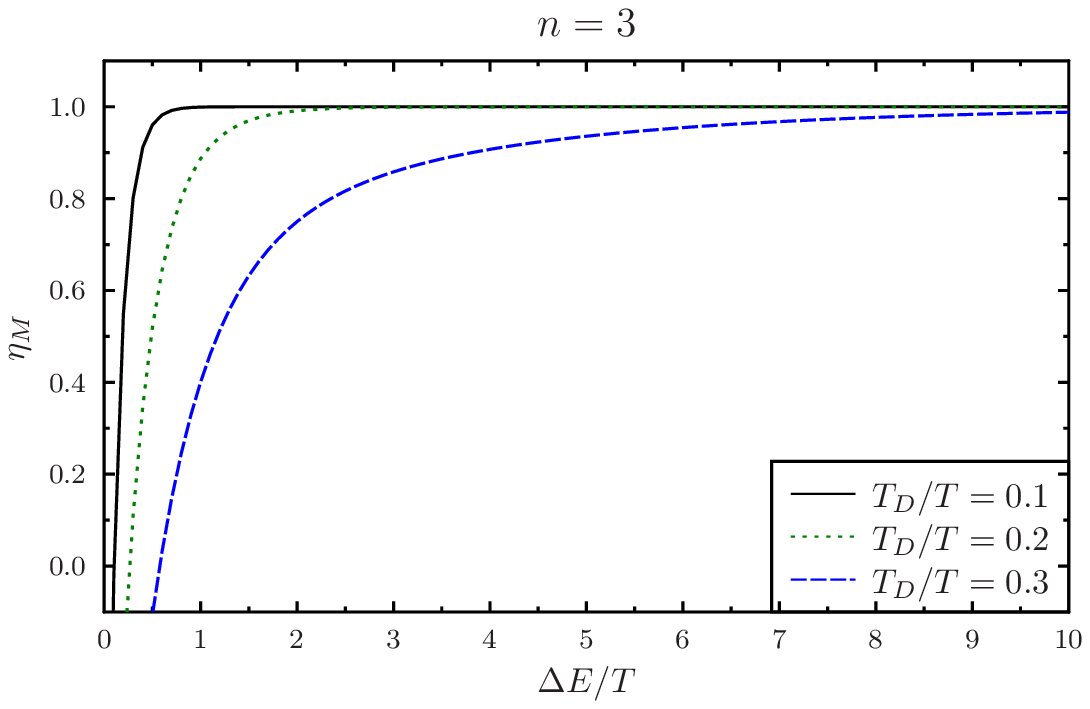}
\includegraphics[width=14cm]{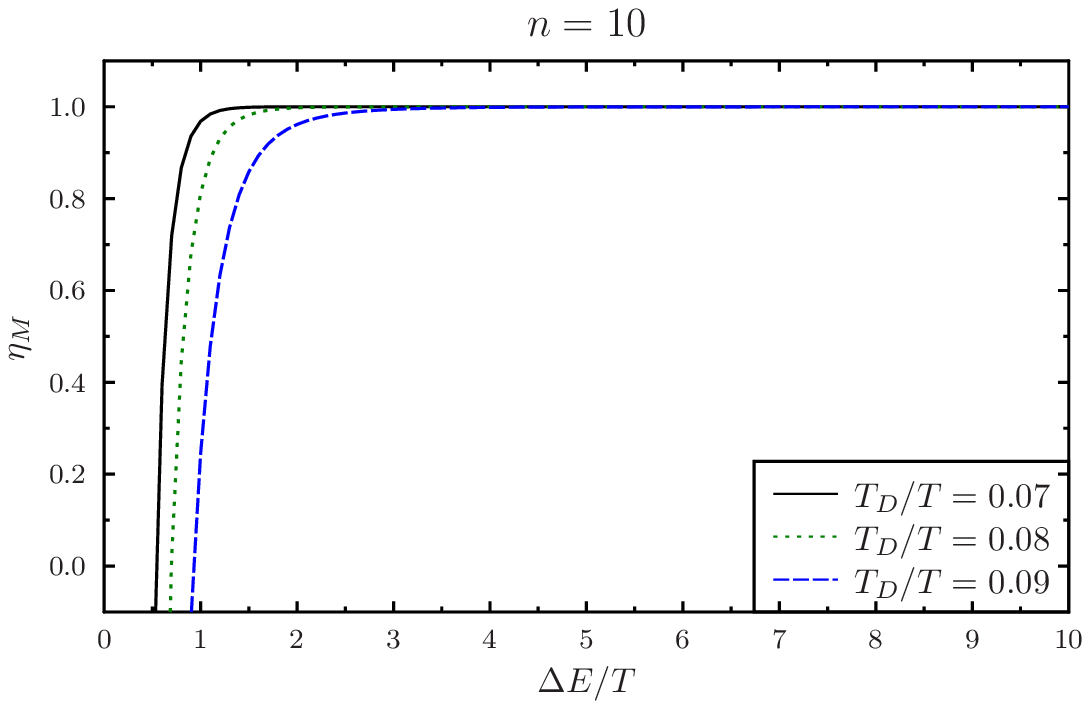}

\caption{Estimated efficiency of machine, $\eta_M$, not taking into account efficiency of demon, for $\sigma_{10} = 1$ and $\Delta \varepsilon = 0$. The efficiency of the machine goes to one if the signal
is error free and the heat produced by the measurement is negligible. The efficiency can become negative since there is a parameter regime in which the machine produces less work than it obtains energy from the demon.\label{fig3}}
\end{figure}

The machine can extract work only from the error-free part of the signal that comes from the demon
\beqn
W= q_n N \Delta E (p_1^{\rm M})^n p_1^{\rm D}  ~,
\eeqn
and so
\beqn
W-W_2 = (q_n-1) N \Delta E (p_1^M)^n p_0^D  - N \Delta E p_1^D  (1 - (p^{\rm M}_1)^n) ~.
\eeqn
Taken together we get
\beqn
\eta_{\rm M} = 1 - \frac{((p_1^M)^{-n}-1)  p_1^D + n \epsilon_n}{(q_n-1) p_0^{\rm D}}~.
\eeqn
The machine's efficiency is not bound by the Carnot limit because it is not a heat engine; it receives
a signal from the demon to do its work.  The
efficiency is shown for some parameter values in figure \ref{fig3}.

Since we do not want to construct a machine that does the demon's work, we
just show that any machine which does this work brings the total efficiency of the
combined system of demon and machine down below the Carnot limit. We do this by
noting that the demon must be able to empty $N$ energy levels, the result of its
computation, to emit the signal
to the machine. This amounts to reducing its entropy and, since the
process is cyclic, that it must
have extracted from its heat bath at least the energy necessary for this reduction.

\begin{figure}[ht]

\includegraphics[width=14cm]{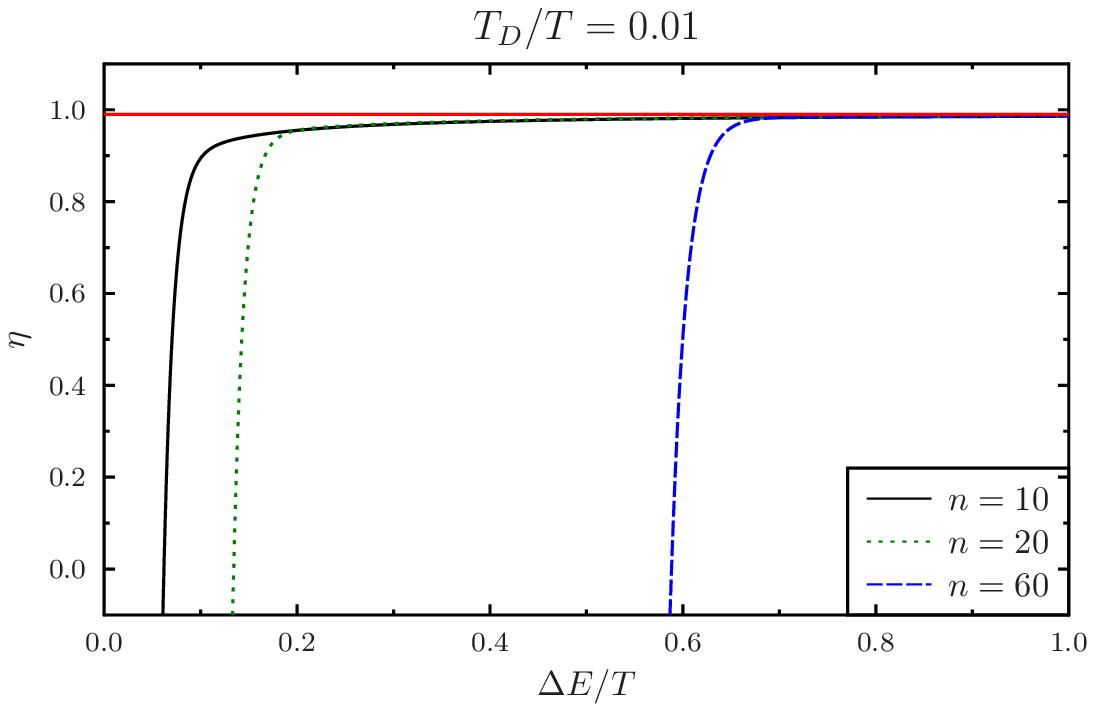}
\includegraphics[width=14cm]{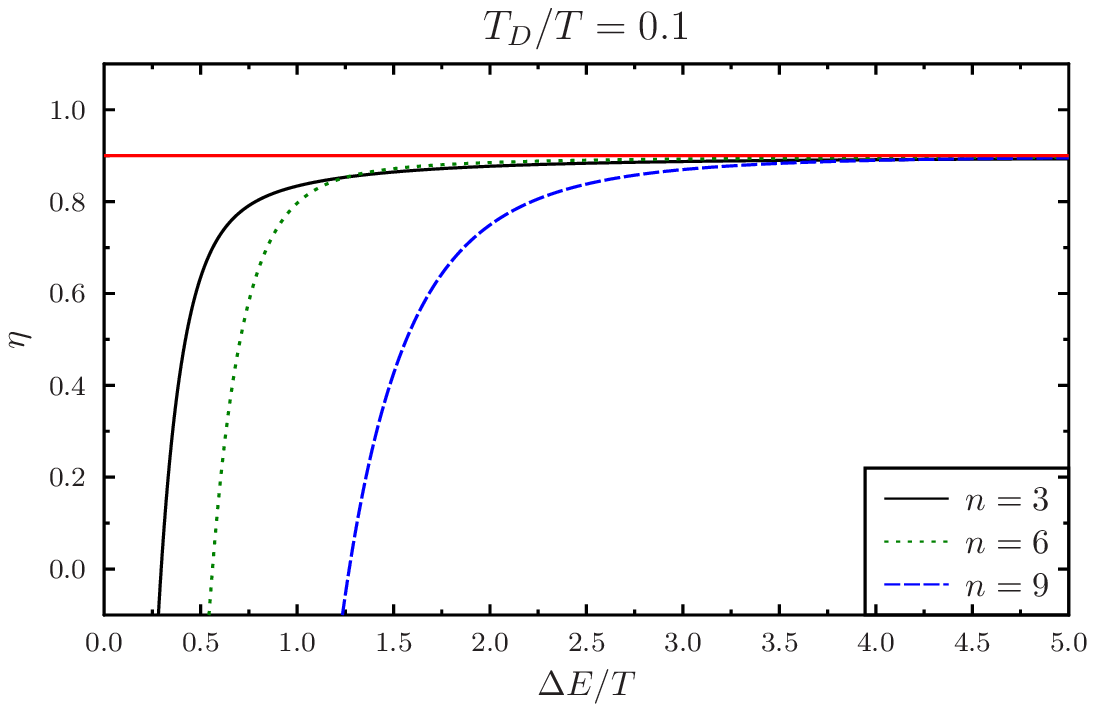}

\caption{Upper limit for efficiency of total system for $\sigma_{10} = 1$ and $\Delta \varepsilon = 0$. The solid (red) horizontal lines indicate the Carnot limit. \label{fig4}}
\end{figure}

To derive an upper bound we use the simplified case in which
 $\sigma_{10} = 1$ and $\Delta \varepsilon = 0$ and the signal is noise-free, because
these additional corrections would only decrease the efficiency.  To further simplify the
estimate we assume that the machine returns to the demon the energy of the signal  that triggered 
the stimulated emission, $W_2$, and the work output is thus only the surplus (again, if this return is imperfect it
would only decrease the efficiency). Since the process
is cyclic and reversible, $\Delta S = Q_2/T_{\rm D}$, where $\Delta S$ is the change in entropy in the demon's two-level system
\beqn
\Delta S &=& - \left( 1 - (p_1^{\rm M})^n \right)  \log \left( 1 - (p_1^{\rm M})^n \right)  +  
 (p_1^{\rm M})^n \log  (p_1^{\rm M})^n ~ \label{entropy}
\eeqn
The total efficiency is then always smaller than
\beqn
\eta \leq \frac{ n (p_1^{\rm M})^n - T_{\rm D} \Delta S/\Delta E }{n(p_1^{\rm M})^n } ~.
\label{eta1}
\eeqn
Now we note that the 2-level distribution with probability $(p_1^{\rm M})^n$ in the upper level that enters the entropy
change (\ref{entropy}) is just that of a thermal distribution with temperature $T/n$ and total
energy  $\Delta E (p_1^{\rm M})^n$, thus \cite{tape}
\beqn
\Delta S \geq \frac{n(p_1^{\rm M}) \Delta E}{T} ~. \label{eta2}
\eeqn
This also leads us to suspect that the system will not work efficiently if the temperature of
the demon $T_{\rm D} \geq T/n$, because it will become very costly to empty the energy
levels. Combining (\ref{eta1}) and (\ref{eta2}) gives 
\beqn
\eta \leq 1- \frac{T_{\rm D}}{T} ~.
\eeqn
 The total efficiency is plotted for some representative values in figure \ref{fig4}.

\section{Discussion}

In the example for the remote demon discussed here the work is produced by stimulated emission in multiples of the incoming energy from demon. This feature is specific to the amplification procedure used and likely
not a general property of the remote demon. As mentioned earlier, if one uses a small potential wall
separating two potential minima of different energies, then the amplification does not necessarily have to
occur in multiples of the incident energy. Alas, it remains to be seen if a concrete example can be constructed
for this case.

\section*{Acknowledgements} 

I thank Ralf Eichhorn, Holger M\"uller and Stefan Scherer for helpful discussion.

\end{document}